\documentclass[
    ,final            
  ]
  {aipproc}

\layoutstyle{8x11double}

\begin{document}

\title{Str\"omgren Photometry of the Delta Scuti Stars 7 Aql and 8 Aql}

\classification{97.10.Sj}
 \keywords      {oscillations, $\delta$
Sct, photometry}

\author{L. Fox Machado}{
 address={Observatorio Astron\'omico Nacional,
Instituto de Astronom\'{\i}a, Universidad Nacional Aut\'onoma de M\'exico,
A.P. 877, Ensenada, BC 22860, Mexico}
}

\author{M. Alvarez}{
  address={Observatorio Astron\'omico Nacional, Instituto de Astronom\'{\i}a, Universidad Nacional Aut\'onoma de M\'exico, A.P. 877, Ensenada, BC 22860, Mexico}
}

\author{L. Parrao}{
  address={Instituto de Astronom\'{\i}a, Universidad Nacional Aut\'onoma de M\'exico, AP 70-264, Mexico, DF 04510, Mexico}
}

\author{J.H. Pe\~na}{
  address={Instituto de Astronom\'{\i}a, Universidad Nacional Aut\'onoma de M\'exico, AP 70-264, Mexico, DF 04510, Mexico}
}

\begin{abstract}
The  preliminary results of new photometric observations of the
$\delta$ Scuti stars 7 Aql and 8 Aql are reported. 51 hr of
$uvby-\beta$ photoelectric photometric data were obtained over the
period June and July 2007 at the San Pedro M\'artir Observatory,
Mexico. Period analyses confirm the three  pulsation modes
discovered in 8 Aql in the framework of the STEPHI 2003 multisite
campaign. For the star 7 Aql we were able to detect only the main
pulsation modes. The standard magnitudes of both stars are obtained.
 The frequency, amplitude and phases of the frequency modes in different
filters are presented.
\end{abstract}

\maketitle


\section{Introduction}

7~Aql (HD~174532, SAO~142696, HIP~92501) is a $\delta$~Scuti
variable discovered in a systematic search and characterization of
new variables in preparation for the COROT mission [1] and was
selected as the main target of the STEPHI XII
multisite campaign in 2003. In this campaign 8~Aql (HD~174589,
SAO~142706, HIP~92524) was used as the only comparison star because
there are no other bright stars in the field-of-view (FOV) of the
4-channel photometer used in the STEPHI network.

\medskip
 Although no check comparison star was implemented in that campaign,
by carefully analyzing the derived
 light curves, differential and non-differential,  it was demonstrated that 8 Aql
is a new $\delta$ Scuti variable. Moreover, it was shown that the
amplitude spectrum of both stars are not superposed. Three and seven
frequency peaks were unambiguously detected with a 99 \% confidence
level  in 8 Aql and 7 Aql respectively and a possible identification
of the observed modes in terms of radial order was performed [2].
CCD photometric observations of 7 Aql and 8 Aql were reported in [3].

\medskip
In the present paper,  preliminary result of new  photoelectric
photometric observations of 7 Aql and 8 Aql are reported.

\section{Observations and data reduction}


The observations were carried out on the nights of June 21, 22, 23, 28, 30 and July 07 and 08
at the Observatorio Astr\'onomico Nacional-San Pedro M\'artir (OAN-SPM), Baja California, Mexico.
The 1.5-m  telescope with the six-channel Str\"omgren spectrophotometer  was implemented.
The observing routine consisted of five 10 s of integration of the
star from which five 10 s of integration of the sky was subtracted.
Two comparison constant stars were observed as well.
The stars were observed for about 51 hr
during the whole campaign (see Table 1).

\begin{table}[!t]\centering
  \setlength{\tabcolsep}{1.0\tabcolsep}
 \caption{Positions, magnitudes and spectral type of the comparison
stars observed with the Str\"omgren spectrophotometer. These data
were taken from the SIMBAD database.} \label{tab:stars_pp}
  \begin{tabular}{cccccc}
\hline
Star&  $ID$ & RA & Dec & V &Sp.T. \\
&&(2000.0)&(2000.0)&(mag)&\\
\hline
 c1 & HD 174046& 18 48 44    &-03 54 01.2  & 9.6&$A0$ \\
 c2 &HD 174625& 18 51 27    &-02 42 05.7  & 9.5&$F5$ \\
\hline
\end{tabular}
\end{table}

\medskip
A set of standard stars was also observed each night to transform
instrumental observations onto the standard system and to correct
for atmospheric extinction.  The instrumental magnitudes ($_{\rm
inst}$) and colours, once corrected from atmospheric extinction,
were transformed to the standard
system ($_{\rm std}$) through the  equations given by [4].

\medskip
The averaged standard magnitudes and indices for 7~Aql and 8~Aql are
listed in Table~\ref{tab:index_pp}. A total amount of 291 data
points were obtained for the former star, while 288 for the latter.
In the case of $H_{\beta}$ 24 measurements were performed for both
stars. The reddening free indices are defined as:
$[m_{1}]=m_{1}+0.18(b-y)$ and $[c_{1}]=c_{1} -0.20(b-y)$ [4].
 Applying the above equations to
the standard stars, an estimation of the uncertainties of each
individual observation was obtained: $\sigma_{v}= 0.011$,
$\sigma_{(b-y)} = 0.006$, $\sigma_{m_{1}}=0.085$,
$\sigma_{c_{1}}=0.015$, $\sigma_{H_{\beta}}=0.015$. The photometric
precision in the instrumental system was: $\sigma_{u}=0.017$,
$\sigma_{v}=0.013$, $\sigma_{b}=0.011$ $\sigma_{y}= 0.009$.

\begin{figure*}[!t]
\includegraphics[width=12cm]{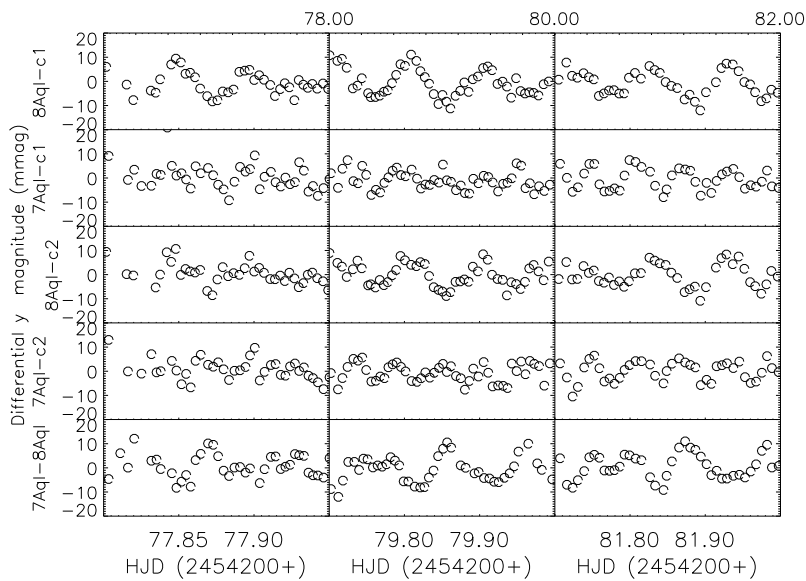}
\caption{Examples of the differential light curves taken with the
Str\"omgren spectrophotometer using the $y$ filter with reference
star HD 174046 $=$ c1 and HD 174625 $=$ c2. The name of each
differential light curve is indicated at left.}
\label{fig:curves_pp}
\end{figure*}

\medskip
Figure~\ref{fig:curves_pp} shows examples of the differential light
curves in $y$ filter of 7 Aql and 8 Aql for three selected nights.

\begin{table}[!t]\centering
 \caption{Averaged standard magnitudes and indices for 7~Aql and 8~Aql.
} \label{tab:index_pp}
  \begin{tabular}{llllllll}
\hline
Star& $V$& $(b-y)$&$m_{1}$& $c_{1}$&$H_{\beta}$&$[m_{1}]$&$[c_{1}]$  \\
&(mag)&(mag)&(mag)&(mag)&(mag)&(mag)&(mag)\\
\hline
7~Aql & 6.894&0.171 &0.180 &0.873&2.755&0.211&0.839  \\
8~Aql & 6.075&0.178 &0.185 &0.822&2.730&0.212&0.786  \\
\hline
\end{tabular}
\end{table}

\begin{figure}[!t]
  \centering
  \includegraphics[width=9cm]{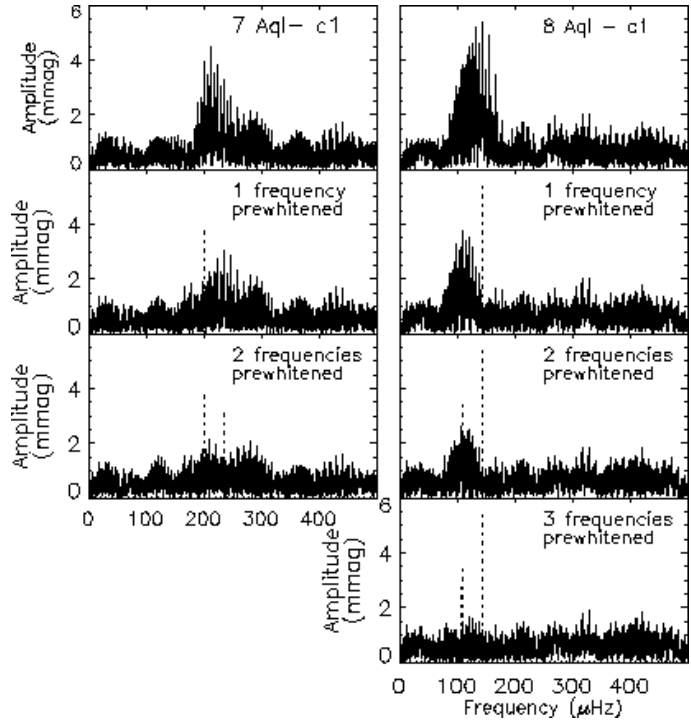}
  \caption{Pre-whitening process in the spectra derived from the
  photoelectric photometric $y$ differential light curves 7 Aql$-$c1 (left), 8 Aql$-$c1 (right).}
  \label{fig:prewh}
\end{figure}

\section{Spectral analysis}
The period analysis has been performed by means of standard Fourier
analysis and least-squares fitting. In particular,  the amplitude
spectra of the differential time series were obtained by means of an
iterative sinus wave fit (ISWF) [5].

\medskip
The amplitude spectra  of the differential $y$ light curves 7
Aql$-$c1, 8 Aql$-$c1  are shown in the top panels of each plot of
Figure~\ref{fig:prewh}.  The subsequent panels of each plot in the
figure, from top to bottom, illustrate the  process of detection of
the frequency peaks in each amplitude spectrum. We followed the same
procedure as explained in [6].

\medskip
The frequencies, amplitudes and phases in the filters $v,b,y$ are
listed in Table~\ref{tab:frec}. The three frequency peaks detected
in the $\delta$ Scuti star 8 Aql in  STEPHI 2003 campaign [2] have
been confirmed in this season. On the other hand,
 only two dominant modes of 7 Aql have been detected.
This is due to the fact that observations from a single site yields a worse windows function.
A non-adiabatic analysis for 7 Aql and 8 Aql using the amplitudes and phases listed
in Table~\ref{tab:frec}
will be given in a forthcoming paper [8].

\begin{table}[!t]\centering
  \setlength{\tabcolsep}{1.0\tabcolsep}
 \caption{Frequency peaks detected in the light curves 7 Aql $-$ c1 and 8 Aql $-$ c1. S/N
is the signal-to-noise ratio in amplitude after the prewhitening process. The formal
error derived from analytical formulae given by [7] are indicated.} \label{tab:frec}
  \begin{tabular}{cccc}
\hline
Freq.&  A & $\varphi$ & $S/N$  \\
($\mu$Hz)&(mmag)&(rad)&\\
\hline
Filter $v$&&&\\
       $200.944 \pm 0.047$ &   $3.85 \pm 0.49$ &   $ -0.27 \pm 0.02$ &  5.1\\
       $236.640 \pm 0.053$ &   $3.12 \pm 0.45$ &  $  +0.32 \pm 0.02$  &  3.6\\
Filter $b$&&&\\
       $200.903 \pm 0.042$ &  $3.69 \pm 0.42$ & $ -0.01 \pm 0.02$  & 5.5\\
       $236.645 \pm 0.058$ &  $2.48 \pm 0.39$ & $  +0.28  \pm 0.02$   & 3.3\\
Filter $y$&&&\\
       $200.878 \pm 0.052$ &   $2.63 \pm 0.37$ & $ +0.21 \pm 0.02$  &  4.7\\
       $236.596 \pm 0.056$ &   $2.28 \pm 0.34$ & $ +0.65 \pm 0.02$  & 3.8\\
\hline
Filter $v$&&&\\
       $143.349 \pm 0.038$ &  $5.43 \pm 0.55$ & $-2.03 \pm 0.02 $ &   9.0 \\
       $110.301  \pm 0.051$ &  $3.65 \pm 0.50$ & $ -0.24 \pm 0.01$&  5.9 \\
       $108.035 \pm 0.058$ &   $2.95 \pm 0.46$ & $ +2.56  \pm 0.01$ &    4.8\\
Filter $b$&&&\\
       $143.316 \pm 0.039$ &  $4.62 \pm 0.49$ & $-1.67 \pm 0.02$ &   9.2\\
       $110.240 \pm 0.037$ &  $2.59  \pm 0.29$ & $-0.09 \pm 0.01$&  4.5\\
       $109.315 \pm 0.060$ &  $3.02  \pm 0.43$ & $-2.83 \pm 0.01$&      5.2\\
Filter $y$&&&\\
       $143.337 \pm 0.031$ & $ 4.02 \pm 0.33$ & $ -1.67 \pm 0.01$&   9.6\\
       $110.230 \pm 0.043$ & $ 2.12 \pm 0.29$ & $ +0.03 \pm 0.01$&   4.8\\
       $109.342 \pm 0.054$ & $ 2.57 \pm 0.31$ & $ -3.03 \pm 0.019$&    5.8\\
\hline
\end{tabular}
\end{table}

\begin{theacknowledgments}
This work has received financial support from the UNAM under grants
PAPIIT  IN108106 and IN114309. Special thanks are given to the technical staff
and night assistant of the San Pedro M\'artir Observatory.  This
research has made use of the SIMBAD database operated at CDS,
Strasbourg (France).

\end{theacknowledgments}



\bibliographystyle{aipprocl} 


\IfFileExists{\jobname.bbl}{}
 {\typeout{}
  \typeout{******************************************This is due to the worse windows function of out}
  \typeout{** Please run "bibtex \jobname" to optain}
  \typeout{** the bibliography and then re-run LaTeX}
  \typeout{** twice to fix the references!}
  \typeout{******************************************}
  \typeout{}
 }



\end{document}